\documentclass[reviewcopy,11pt]{elsart}
\usepackage[dvips]{graphicx}
\usepackage{epsfig}
\usepackage{amsmath}
\usepackage{amssymb}
\usepackage{subfigure}

\begin{document}
\begin{frontmatter}

\title{Percolation and roughness in ballistic deposition of complex
shapes}

\author[label1]{Linder C. DaSilva}
\author[label1]{and Marconi S. Barbosa\corauthref{cor1}}
\author[label2]{and Osvaldo N. Oliveira Jr.}
\author[label1]{and Luciano da F. Costa}
\corauth[cor1]{FAX: 55 16 273 9879, {\it email:} marconi@if.sc.usp.br}
\address[label1]{Cybernetic Vision Research Group Instituto de
F\'{\i}sica de
  S\~ao Carlos University of S\~ao Paulo 13560-970 S\~ao Carlos, SP,
Brazil}
\address[label2]{Polymer Research Group, DFCM-IFSC, Universidade de S\~
ao
  Paulo, S\~{a}o Carlos, SP, Caixa Postal 369, 13560-970, Brasil}

\begin{abstract}
  This paper investigates how the measurement of geometrical features of
structures obtained from ballistic deposition of objects with complex
shapes, particularly neuronal cells, can be used for characterization
and  analysis of the shapes involved.  The   experiments were performed
on both synthetic (prototypes) and natural shapes. Two measurements were
considered, namely the surface roughness and the critical percolation
density, with the former providing better discrimination of shape
characteristics for alpha and beta neuronal cells of cat retina.

\end{abstract}

\end{frontmatter}

\section{Introduction}

Statistical mechanics has been extremely successful in connecting the
emerging properties of a system of many elements to the microscopic
description of its basic components. Because several natural systems
involve basic elements with very simple geometry (point particles, for
instance), relatively little attention has been given to systems
containing elements with more complex geometry. Yet, understanding the
statistics of complex shape elements is essential for a number of
phenomena, especially in biology. A prototypical such situation is the
central nervous system of animals, where a large number of neuronal
cells, each with its intricate and particular shape, interconnect one
another to produce emergent behavior. Indeed, neuronal cells are among
the natural objects with the most complex shapes, which are required to
make selective connections with close and distant targets
\cite{Kandel:1996}. The functional properties of a mature neuronal
system are mainly determined by its synaptic weights, but the connecting
pattern is ultimately a consequence of spatial interactions during the
development of the system. Therefore, it is important to devise means to
characterize the variety of neuronal shapes found in living beings.  In
addition, such morphological characterization can be used for
classification in taxonomical studies (e.g.
\cite{Costa:1999,Costa:2004,Barbosa:2003,Jones:2001}) and for diagnosis
of pathologies depending on neuronal shape alterations (e.g.
\cite{Costa:2002,Barbosa:2004}).

Several measurements have been proposed to characterize the morphology
of complex shapes
(e.g.~\cite{Costa:2001b,Costa:2002,Montague:1991,Costa:2001}), but the
analysis of complex aggregates found in Nature should also take into
account the potential for interaction and contact between the
constituent elements. Although several indirect measurements of the
potential for contact between objects --- including the ratio between
the squared perimeter and area of the basic elements, the fractal
dimension \cite{Montague:1991,Costa:2001,Costa:1999} and lacunarity
\cite{Mandelbrot:1983,Einstein:1998,Rodrigues:2004} --- have been tried
in the literature, only recently a more direct approach to the
quantification of the connectivity potential between objects has been
suggested \cite{Costa:2003,Costa:2004}. A comparative discussion between
percolation-based approaches and other more indirect measurements of
cell morphology can be found in~\cite{Costa:2003}. In the latter work,
percolation was reached by establishing a connecting path between the
left and right sides of a working space square, with the progressive
addition of shapes to the square. In the case of neuronal cells, a
connection was established during the simulation whenever a point of the
dendrite was found to overlap a point of an axon.  It has been suggested
\cite{Costa:2003} that the critical density of elements characterizing
percolation provides one of the most direct indications of the potential
of those elements to form connected components.  In case a single type
of shape is used in the percolation simulations, it is possible to
employ statistical mechanics to associate the intrinsic geometrical
properties of the shape under analysis with the size of the emerging
aggregates.

The present article investigates how the geometrical features of
elements obtained with ballistic deposition
\cite{Barabasi:1995,Sutherland:1966,Vold:1959} can be used to predict
their potential for contact with each other. To our knowledge, this is
the first use of ballistic deposition of elements assumed to have
complex shapes for analysis of geometrical properties. Ballistic
deposition is justified as an alternative model for characterizing
neuronal (or any other shape) connectivity because it contemplates the
biological situation where the growing neurites of migrating neuroblasts
connect at the first contact with another neuronal cell. This feature is
complementary to the approach reported in~\cite{Costa:2003}, in which
the neurons are stamped one over the other, and to the more recent
investigation of ~\cite{Regina:cond-mat}, where the neurons are allowed
to grow and percolate. Two measurements are considered to evaluate
connectivity: the roughness of the aggregates surfaces and the critical
percolation density.

The article starts by presenting the methodology adopted to simulate the
ballistic deposition of aggregates and to obtain the corresponding
measurements, after which the results are presented and discussed.

\section{Methodology}

\begin{figure}[h]
\centering \includegraphics[scale=0.36]{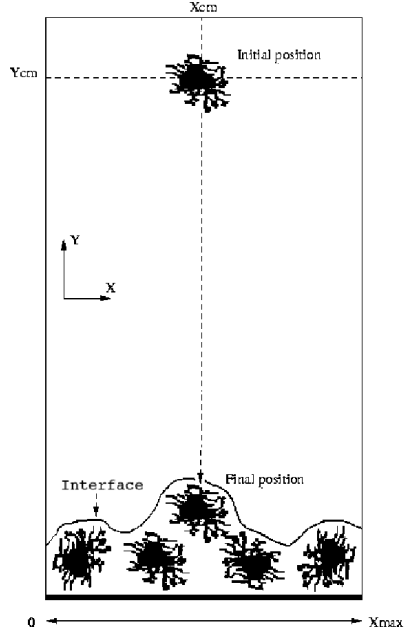}
\caption{Schematic representation of the systems used in simulations.
Deposition is simulated on a rectangular grid with the surface at the
bottom \label{fig:lattice}}
\end{figure}

In the standard on-lattice ballistic deposition model
\cite{Barabasi:1995}, a point particle is dropped from a randomly chosen
position above the surface, located at a distance larger than the
maximum height of the interface (the top surface of the aggregate in
Figure~\ref{fig:lattice}). The particle follows a straight vertical
trajectory until it reaches the surface of the aggregate, whereupon it
sticks.  The standard model has been extended to support the deposition
of simple convex objects such as disks and spheres
\cite{Meakin:1988,Viot:1993,Barabasi:1995}.  Such studies aimed
essentially at describing the scaling properties of growing surface and
not at characterizing the objects being deposited. The model developed
here is an extension of traditional ballistic approaches, by considering
the deposition of generic planar objects, such as neuronal cells, and
investigating how geometrical features extracted from the aggregates --
namely the roughness of the aggregate surface and the critical
percolation density - can be used to characterize the shapes.

The surface roughness is a measure of fluctuation in the height of the
interface which can be defined as follows:

\begin{equation}
r(w,t) = \sqrt{\frac{1}{L}\sum_{i=1}^{L}[h(i,t)-\bar{h}(t)]^2}
\end{equation}

Where $\bar{h}(t)$ is the mean height of the aggregate surface and
$h(i,t)$ is the height of the column $i$ of the lattice at time $t$.
Note that the roughness is a function of time.  The height values were
obtained by thresholding the lattice image after applying the Distance
Transform algorithm \cite{Cuisenaire,Costa:2001b}.

Percolation is a concept from statistical mechanics which corresponds to
a phase transition in the system, as in the case of a cluster of
deposited objects extending throughout the lattice connecting two
lateral borders. This phenomenon is characterized in our simulation by
the number of deposited objects at the percolation time.

In this study we considered 5 artificial planar objects and 61 planar
neuronal cells, where the latter were normalized with respect to their
diameter to minimize the effects from the different sizes of the
objects. The artificial objects include several shapes to illustrate the
percolation approach, while the natural objects refer to real neuronal
cells from the cat retina.

Four data structures were used during the simulations: a matrix ({\tt
lattice}) 6000x1000 to store the objects, a vector ({\tt surface}) to
store the current height of each {\tt lattice} column, and two lists
({\tt flood\_points} and {\tt shape\_points}), respectively, to check
for percolation and to store the points of the current object being
deposited.

The algorithm includes five main modules repeated in a loop, which are
performed over the image points stored in {\tt shape\_points}:

\begin{enumerate}

\item Each object initiates its trajectory after having its center of
mass to coincide, see Figure~\ref{fig:lattice}, with the position
(${x_{cm},y_{cm}}$), where $y_{cm}$
is the maximum height of the {\tt lattice} and $x_{cm}$ is generated
from high-quality pseudo-random-number generator giving a uniform
distribution between 0 and $x_{max}$.

\item The object is rotated between 0 and {$2\pi$}, around its center of
mass.

\item In order to make it possible for the deposition of the object to
occur, the {\tt surface} is checked to determine the contact point. This
is followed by updating the vector {\tt lattice}, which stores every
deposited object.

\item The roughness is computed.

\item The {\tt lattice} is checked for percolation by running a flooding
algorithm which tries to find a connected path from the left side of the
lattice to the right side. If a connected path is found, the percolation
has occurred; the number of deposited objects is recorded and execution
is finished.  Otherwise, the entire loop is repeated.

\end{enumerate}

Two hundred simulations were performed for each case in order to compute
the average percolation. All results have been obtained using an
Openmosix cluster consisting of 10 Pentium IV 2.8 GHz and code developed
in C++; each simulation needed about 15 hours of processor time to
complete.

\section{Results and Discussion}

\begin{figure}[ht]
  \begin{center}
    \includegraphics[scale=.23,angle=0]{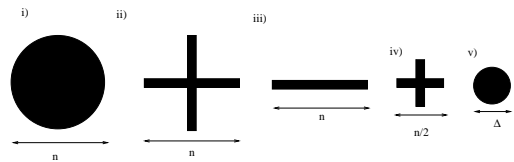}
  \caption{Prototypical generic shapes used in the ballistic deposition
computer    experiment. The diameter $\Delta$ in the last shape of this
figure is defined by the constrained area of the circle which should
match n pixels,
    i.e.  $\Delta=2\sqrt{n/\pi}$~\label{fig:artificial}}
  \end{center}
\end{figure}

The procedure adopted to simulate ballistic deposition and assess
connectivity is initially applied to the artificial shapes presented in
Figure~\ref{fig:artificial}. These forms were chosen to illustrate how
the gradual occupation of the Euclidean plane takes place: i) a ball
with diameter of $n$ pixels; ii) a cross with diameter of $n$ pixels;
iii) a line segment with $n$ pixels; iv) a cross with area of $n$ pixels
(diameter $n/2$); v) a ball with area of $n$ pixels. An important issue
is to identify the most important geometrical characteristic for
percolation with ballistic deposition, whether it is the linear
extension or the area. The analysis of percolation threshold should give
an overall view of what to expect in cases with more general forms.

\begin{figure*}[htb]
  \begin{center}
    \includegraphics[scale=.18,angle=0]{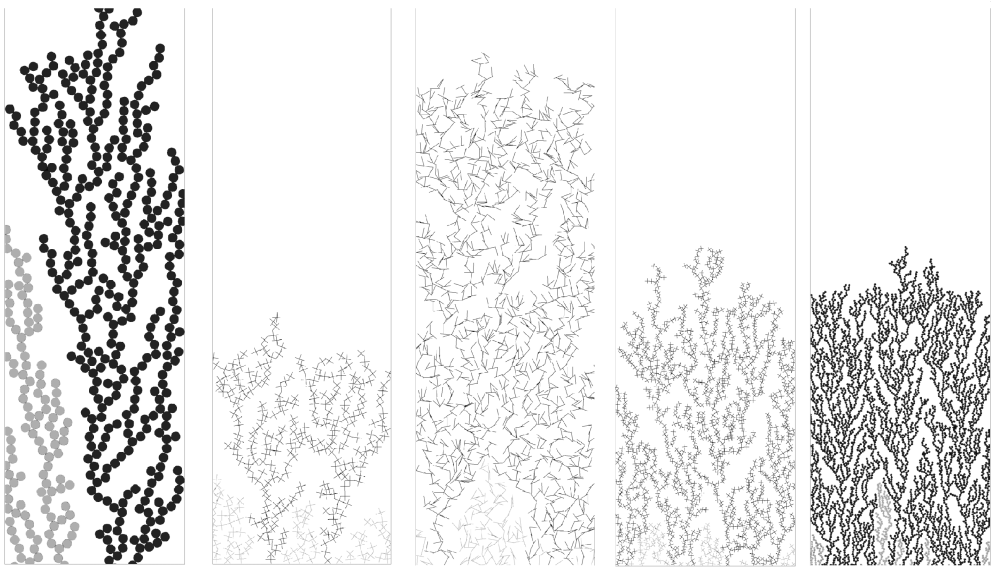}
  \caption{A realization of the ballistic deposition percolation of the
prototypical shapes considered in Figure~\ref{fig:artificial}. Connected
groups in darker color correspond to the percolated
cluster.~\label{fig:art:choral}}
  \end{center}
\end{figure*}

\begin{figure*}[htb]
  \begin{center}
    \begin{tabular}{ccc}
\subfigure[]{\includegraphics[scale=.17,angle=-90]{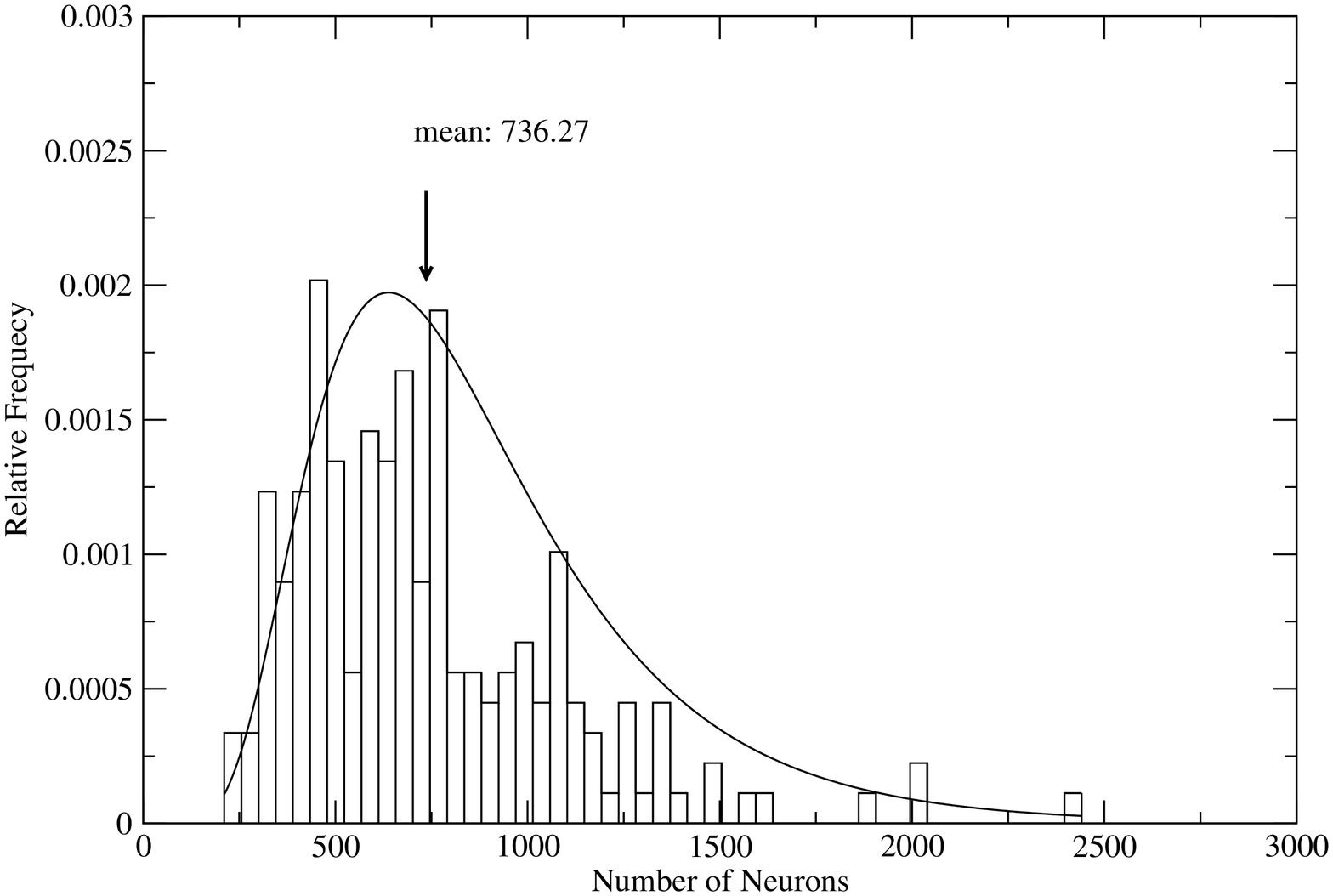}}&
\subfigure[]{\includegraphics[scale=.17,angle=-90]{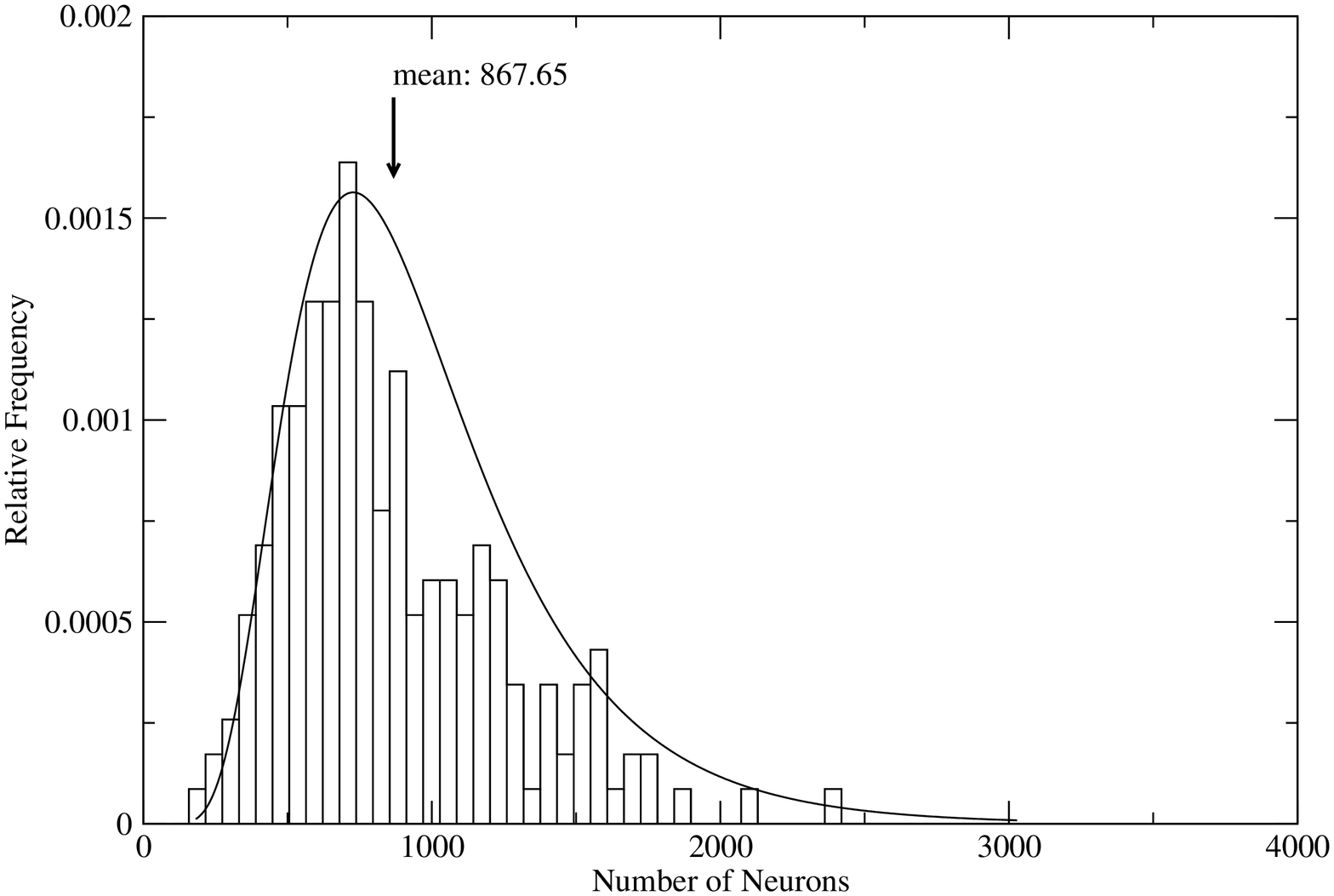}}&
\subfigure[]{\includegraphics[scale=.17,angle=-90]{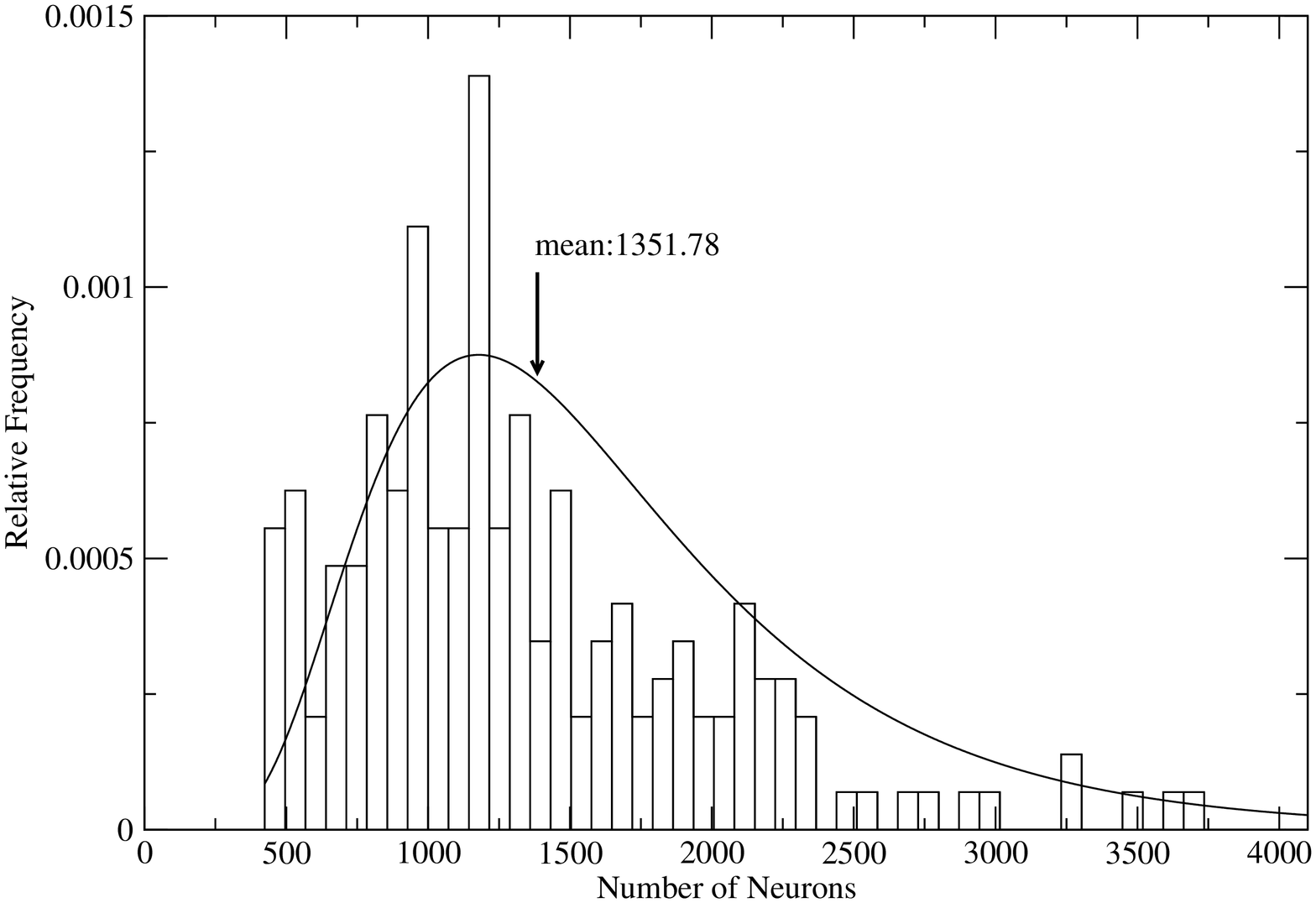}}\\
\subfigure[]{\includegraphics[scale=.17,angle=-90]{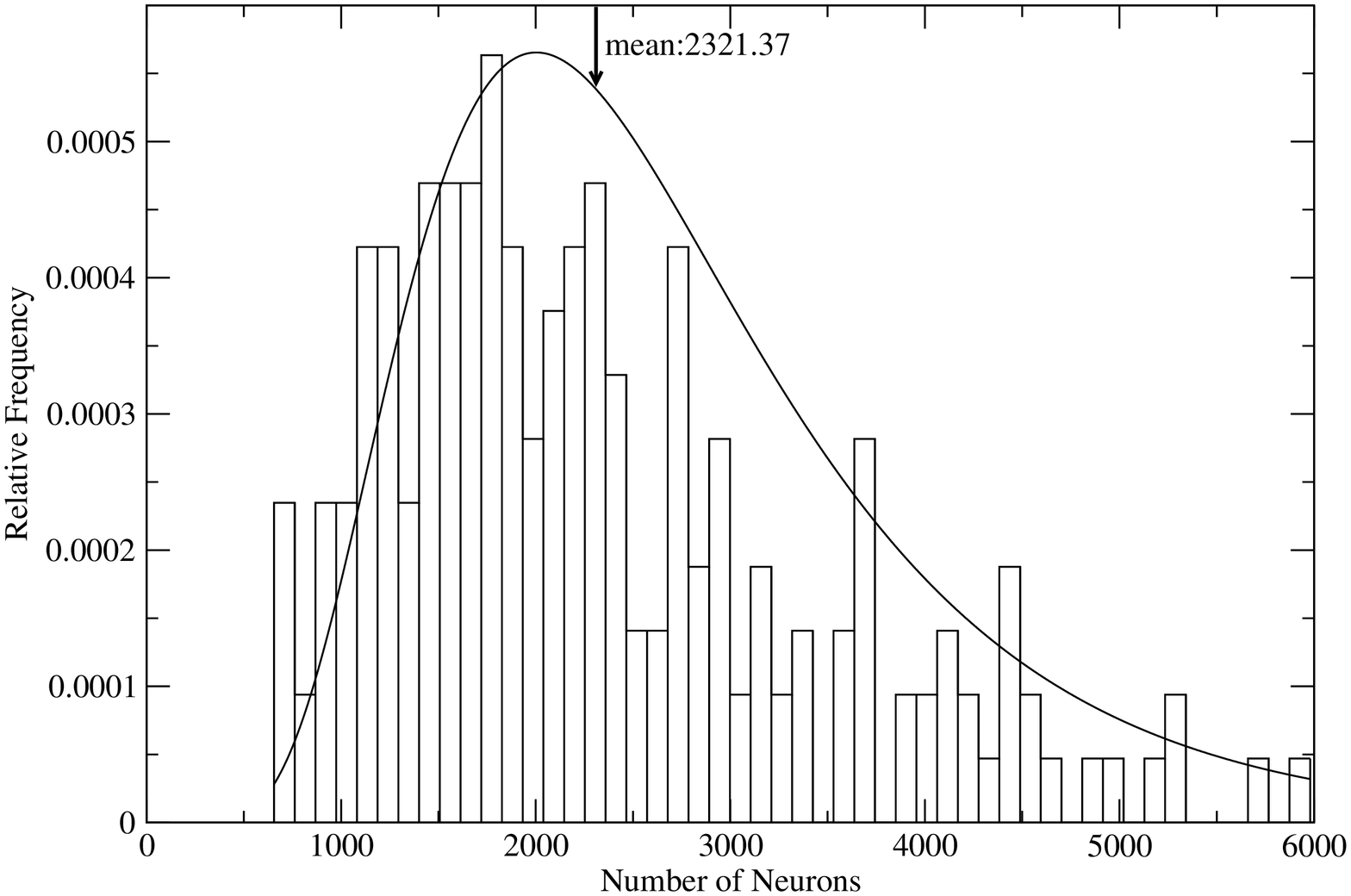}}&
\subfigure[]{\includegraphics[scale=.17,angle=-90]{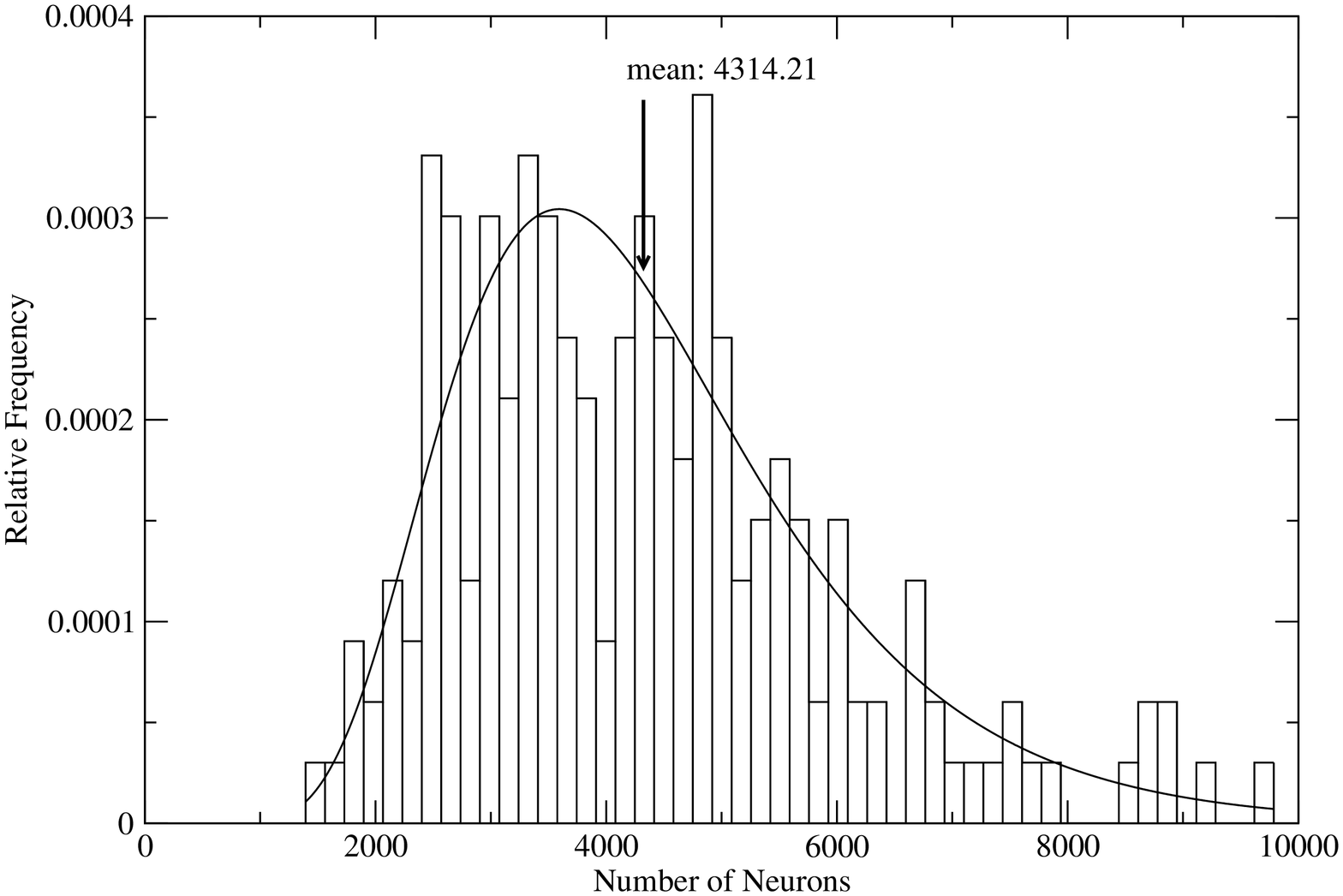}}&
    \end{tabular}
    \caption{Histograms from the computer simulation of the number of
neurons required to achieve percolation for the prototypical chorals
illustrated in Figure~\ref{fig:art:choral}.~\label{fig:art:hist}}
\end{center}
\end{figure*}

Figure~\ref{fig:art:choral} shows an example (realisation) of typical
simulated chorals for each of the artificial shapes.
Figure~\ref{fig:art:hist} shows the histogram of the critical number of
elements to achieve percolation for two hundred simulations for each
shape.

\begin{figure*}[htb]
  \begin{center}
    \includegraphics[scale=.35,angle=-90]{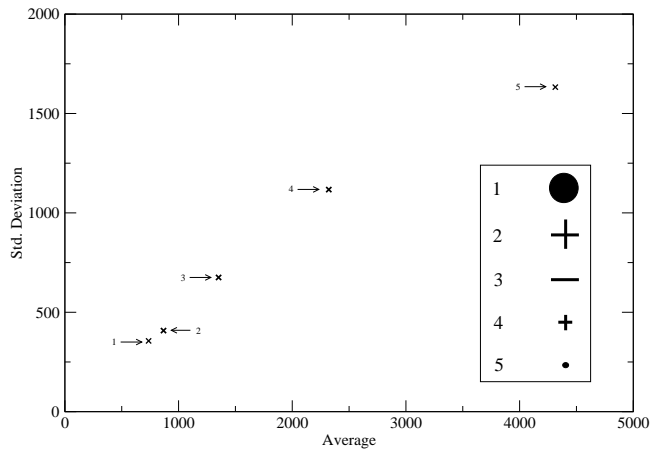}
  \caption{A scatter plot based on the mean and standard deviation of
the histograms of Figure~\ref{fig:art:hist}, showing the influence of
the geometry on percolation times.~\label{fig:art:scatter}}
  \end{center}
\end{figure*}

These histograms were fitted to a log-normal distribution curve, from
which we extracted the global scalar features (e.g., the mean, the
standard deviation, the maximum position, etc.).  The first two of these
measures were used to produce the scatter plot of
Figure~\ref{fig:art:scatter}, which indicates that the morphology of the
individual shapes plays an important role in determining the percolation
threshold. As one could expect from the shapes investigated, the
critical density varies considerably, with the large ball leading to
percolation with the smallest number of objects (i.e. lowest density),
while the highest critical density was obtained with the small ball.
Note also that the average and the standard deviation of the critical
densities are strongly correlated. As will be shown later, this
correlation is maintained for more complex shapes.

\begin{figure*}[htb]
  \begin{center}
    \includegraphics[scale=1.0,angle=0]{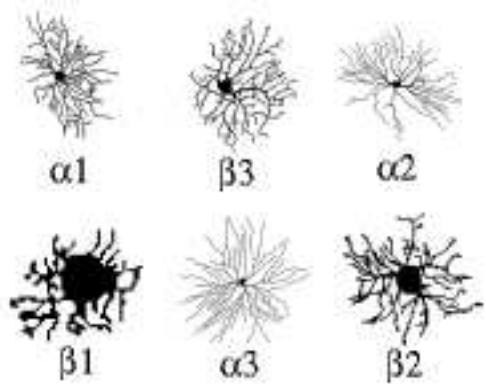}
  \caption{Examples of the alpha and beta morphological classes of cat
ganglion neuronal cells, three of each. ~\label{fig:neuron}}
  \end{center}
\end{figure*}

Having observed a correlation between percolation threshold and shape
for the artificial elements, we now proceed with the ballistic
deposition of neuronal ganglion cells belonging to two of the main
morphological/physiological classes of the domestic cat retina (alpha
and beta classes). These are more complex shapes exhibiting limited, but
definite, fractality ~\cite{Costa:1999}. Figure~\ref{fig:neuron}
illustrates a few examples of alpha and beta cells. There are a great
variety of geometry characteristics among these cells even inside each
class. In this paper we use the chorals grown by ballistic deposition as
a toy model for the complex architecture of the actual cells in a mature
network. The influence of cell shape at facilitating connectivity and on
the surface properties is then directly associated with the potential
for percolation.

\begin{figure*}[htb]
  \begin{center}
\subfigure[]{\includegraphics[scale=.3,angle=0]{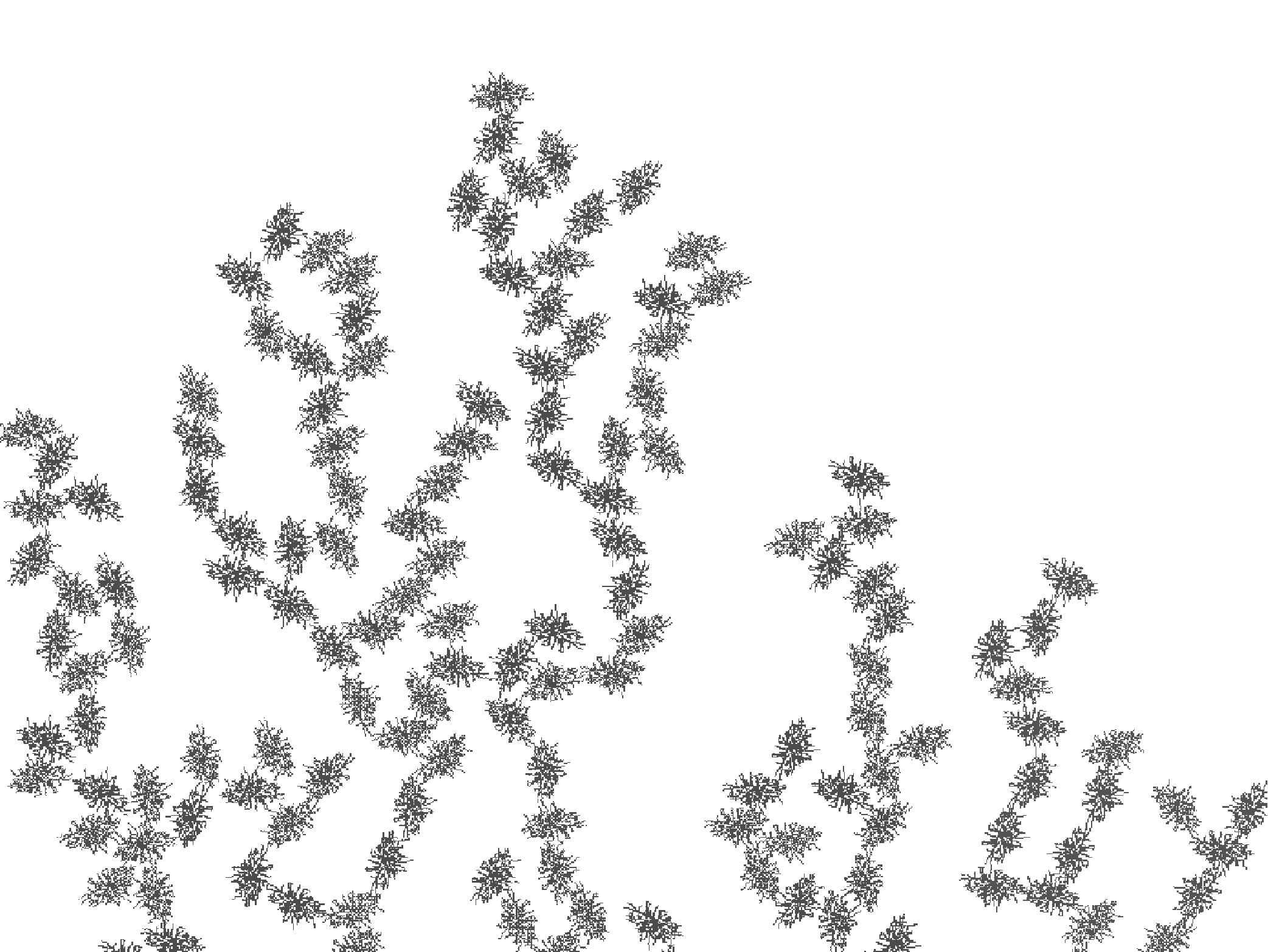}}\\

\subfigure[]{\includegraphics[scale=.3,angle=0]{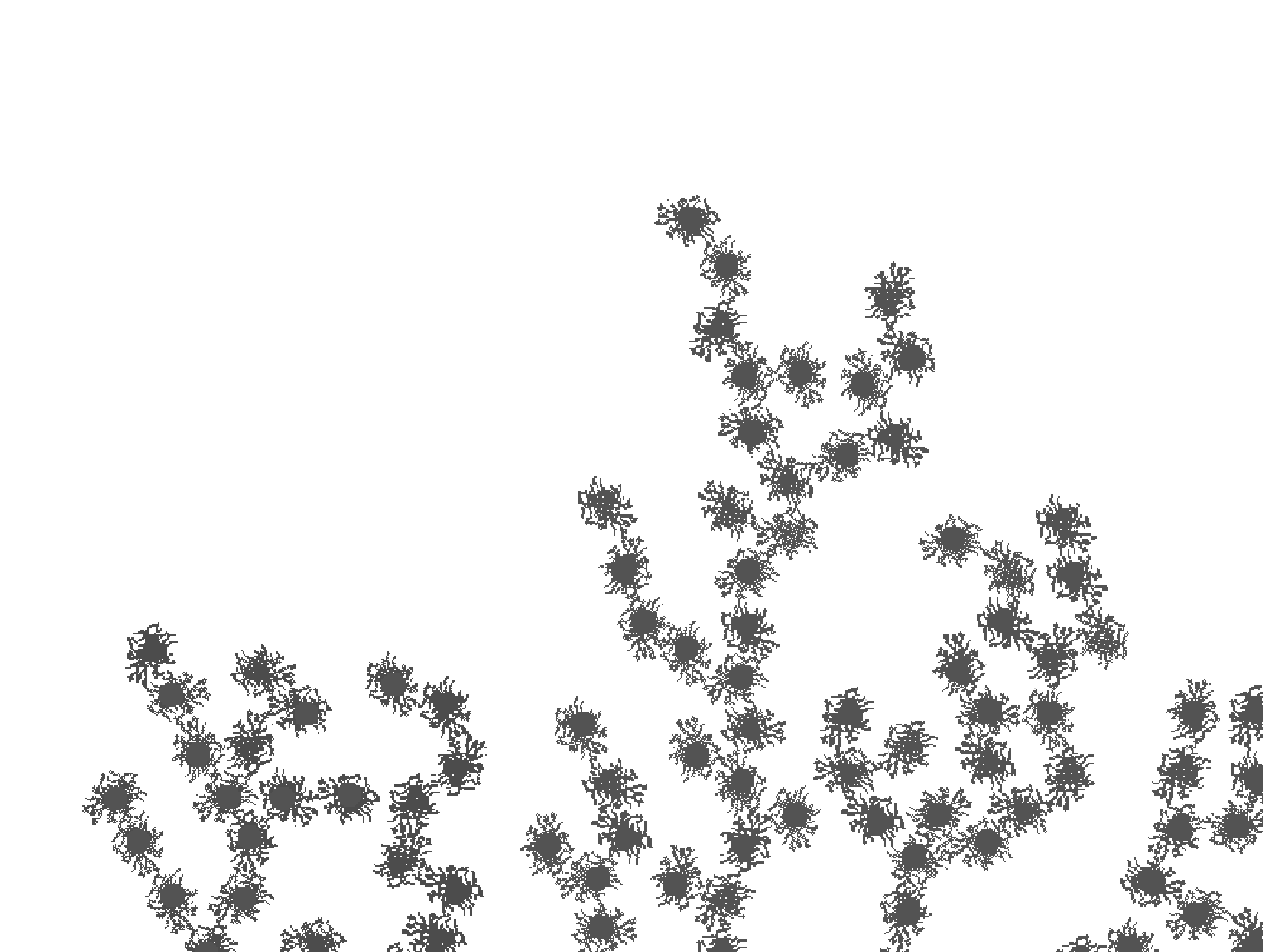}}
    \caption{Two examples of choral formation for representative
examples of neuronal cells, one of each morphological
class.~\label{fig:choral}}\end{center}
\end{figure*}

\begin{figure*}[htb]
  \begin{center}
    \begin{tabular}{ccc}
\subfigure[]{\includegraphics[scale=.17,angle=-90]{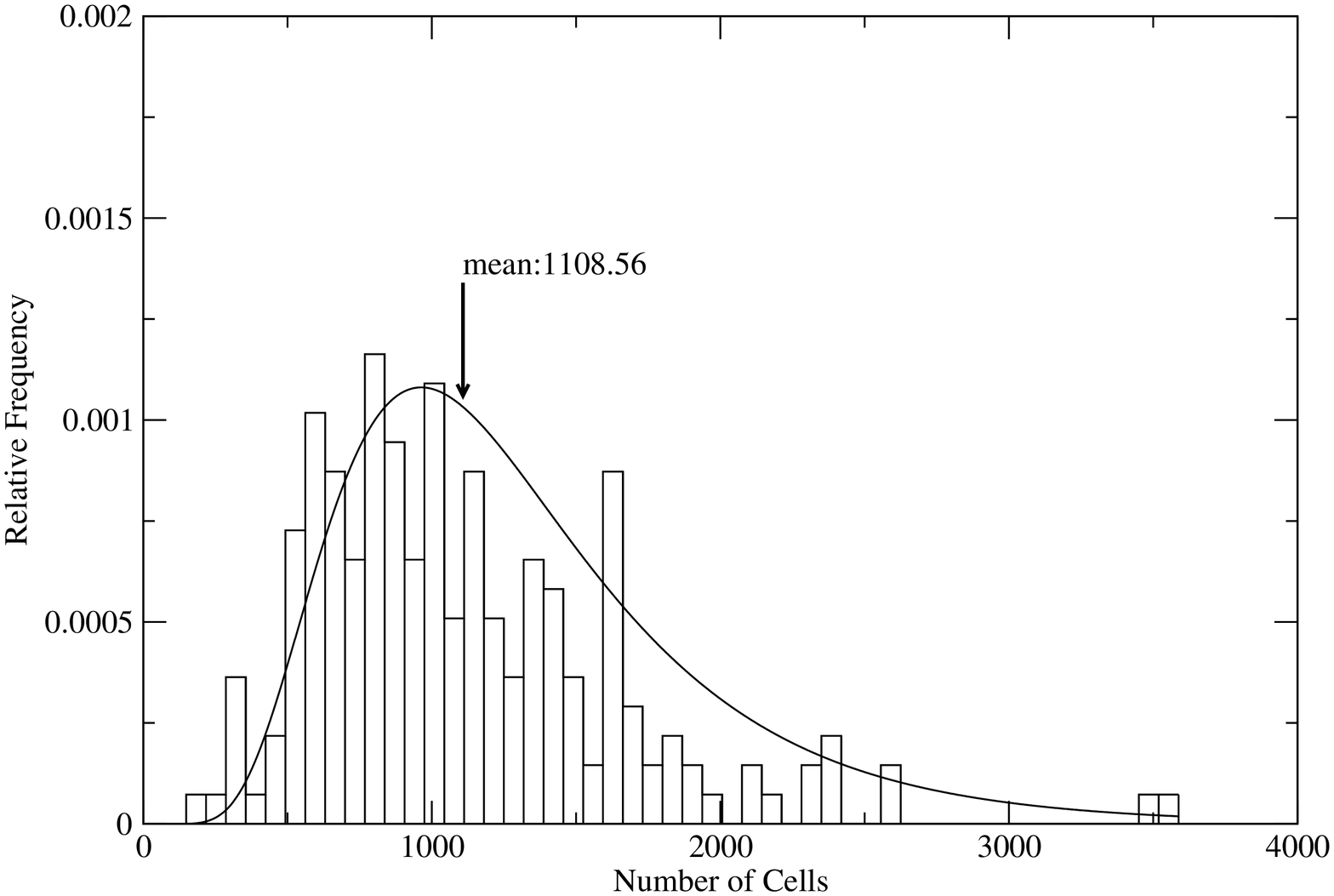}}&
\subfigure[]{\includegraphics[scale=.17,angle=-90]{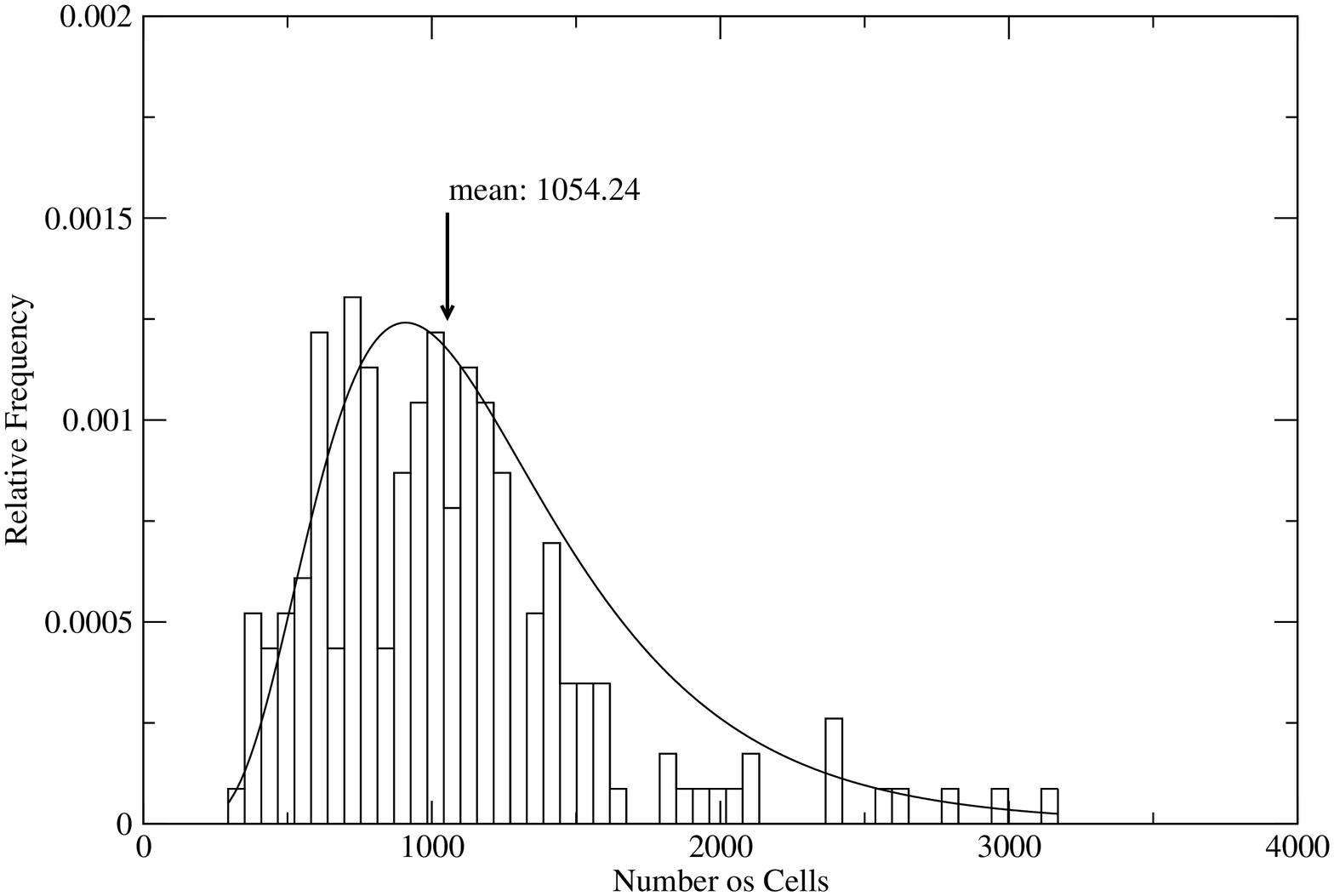}}&
\subfigure[]{\includegraphics[scale=.17,angle=-90]{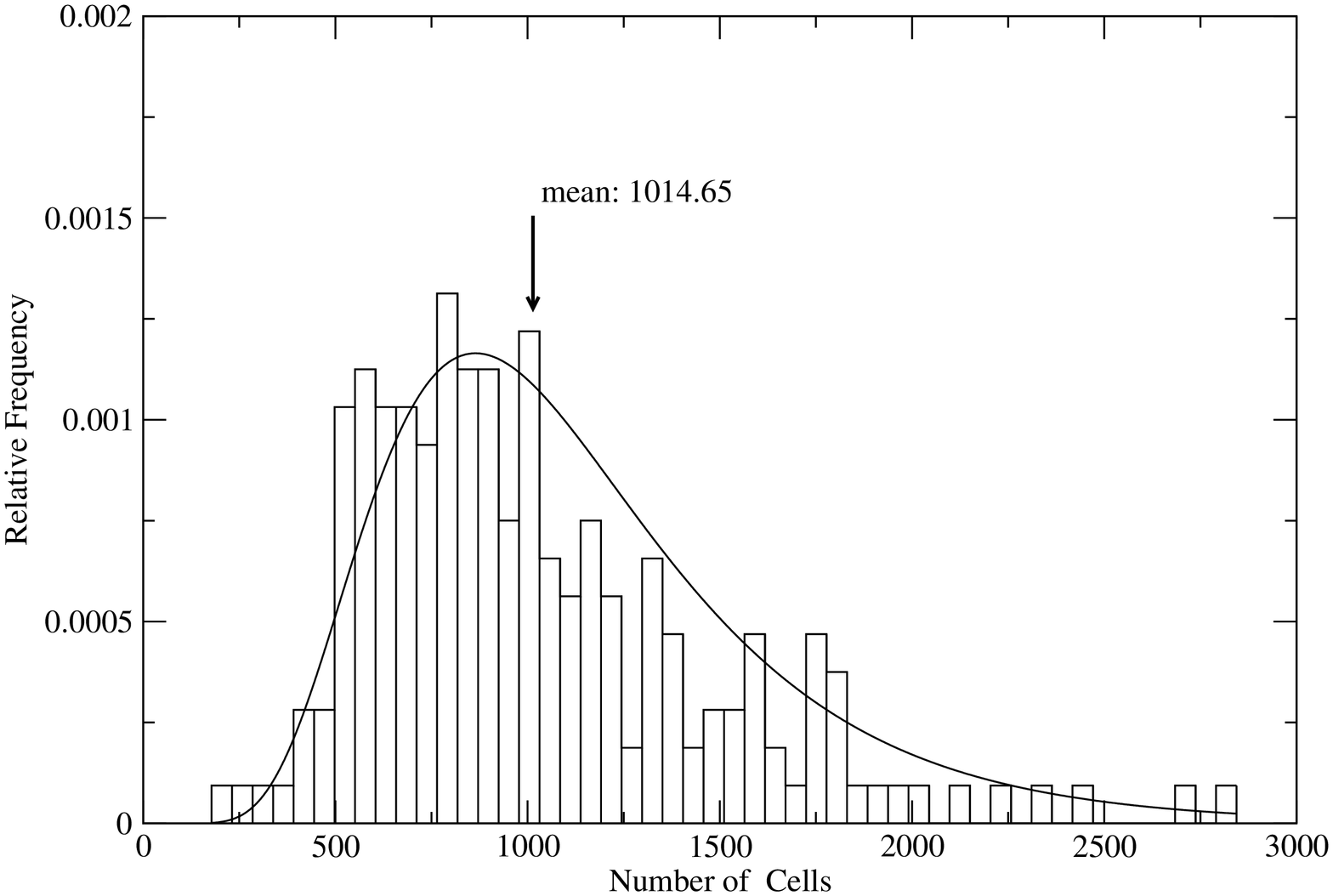}}\\
\subfigure[]{\includegraphics[scale=.17,angle=-90]{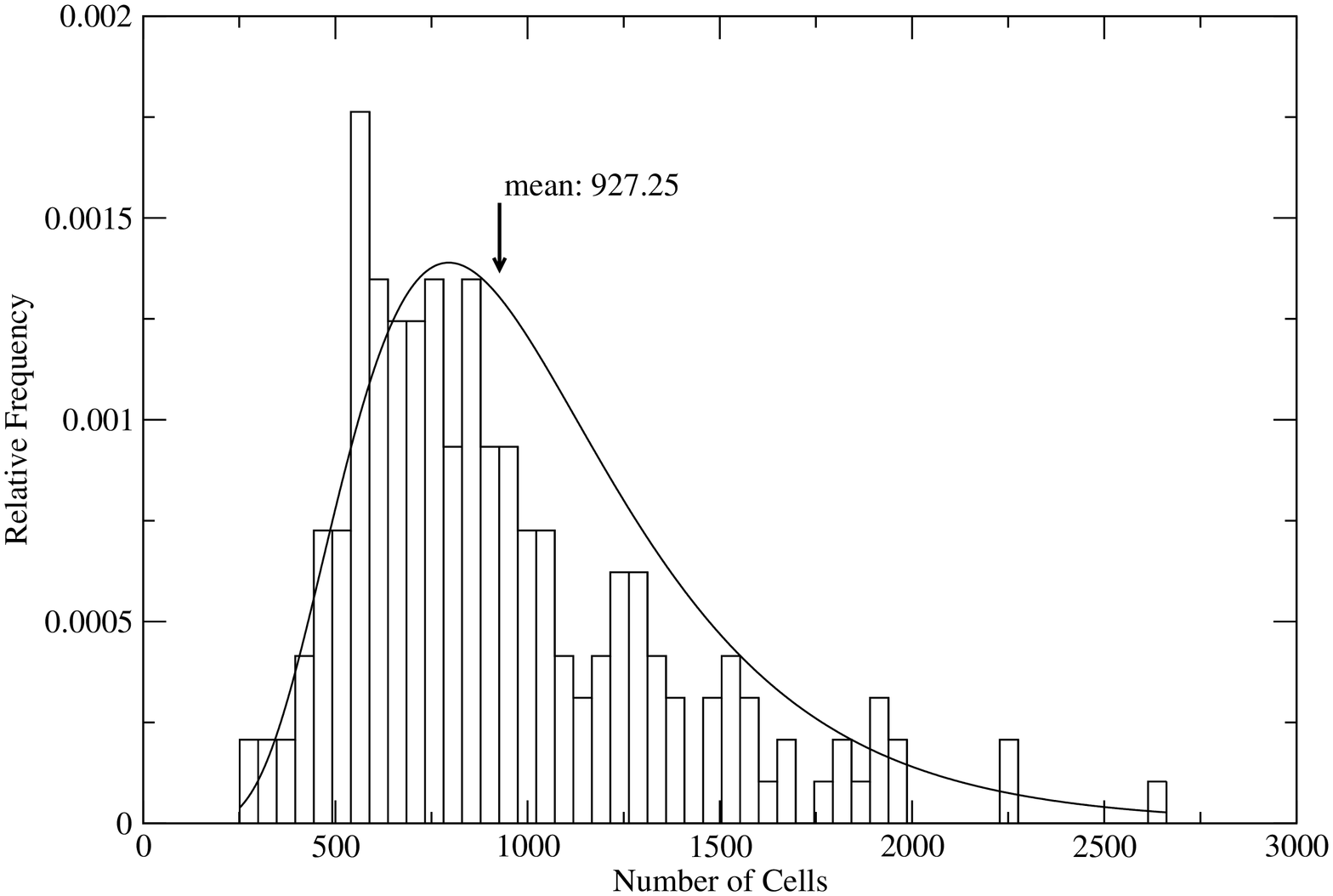}}&
\subfigure[]{\includegraphics[scale=.17,angle=-90]{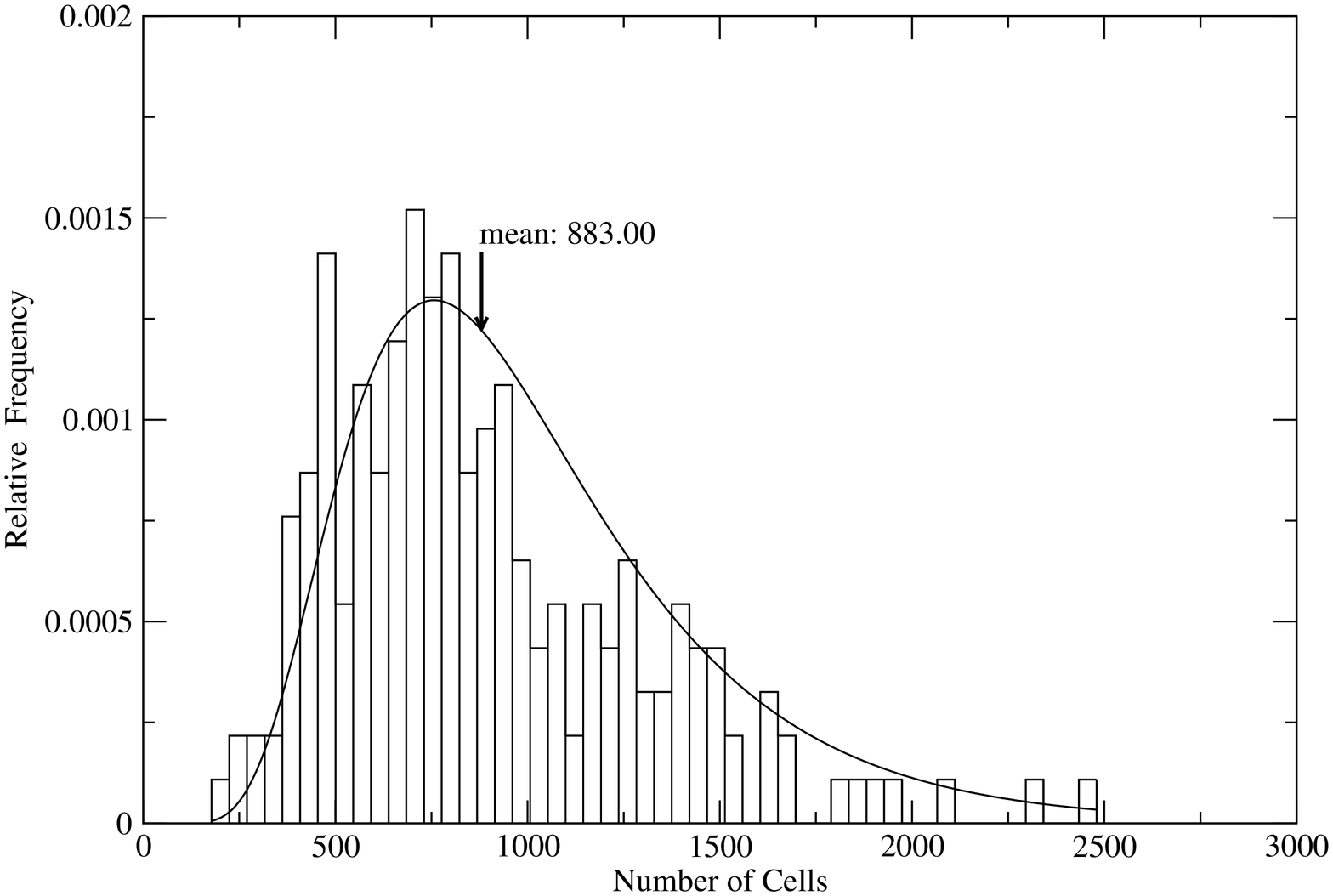}}&
\subfigure[]{\includegraphics[scale=.17,angle=-90]{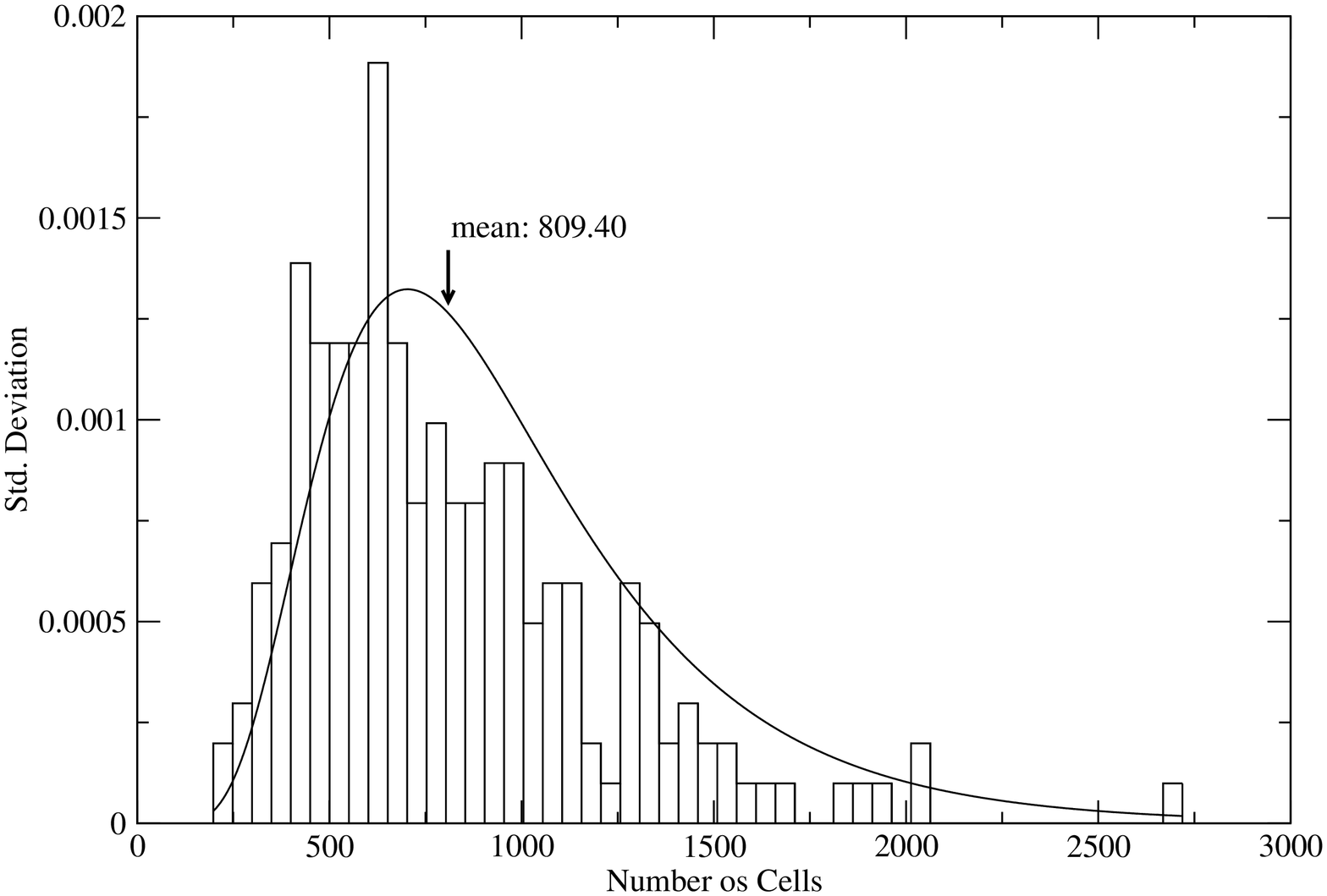}}
    \end{tabular}
    \caption{Histograms from computer simulation for the six example
cells shown in Figure~\ref{fig:neuron}, three of each morphological
type.~\label{fig:hist}}
\end{center}
\end{figure*}

Figure~\ref{fig:choral} shows two examples of chorals obtained by
ballistic deposition of an alpha and a beta cell displayed in
Figure~\ref{fig:neuron}. As we did for the set of artificial cells, a
sequence of two hundred simulations were performed for each of the cells
in our database, with each simulation considering only a single cell.
Histograms for the cells shown in Figure ~\ref{fig:neuron}are displayed
in Figure~\ref{fig:hist}.  The average in each case, indicated by an
arrow, tends to decrease from (a) to (f), indicating a clear decrease of
the potential for connection of those cells.

\begin{figure*}[htb]
  \begin{center}
    \includegraphics[scale=.45,angle=-90]{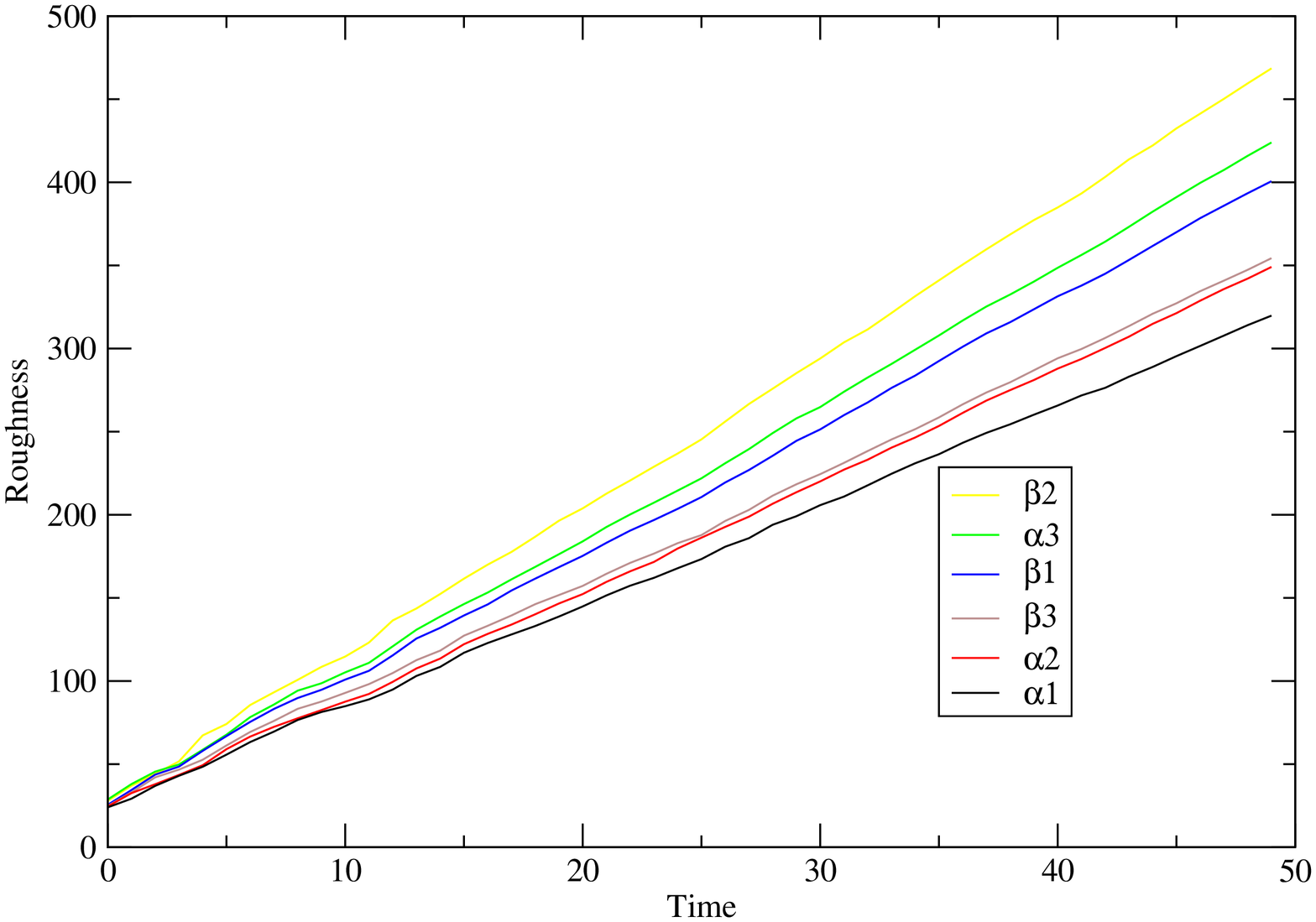}
  \caption{The roughness of the aggregates for a few example cells as a
function of the deposition time, showing the evolution of the
sensitivity of this measure.~\label{fig:rugo}}
  \end{center}
\end{figure*}

\begin{figure*}[htb]
  \begin{center}
    \includegraphics[scale=.45,angle=-90]{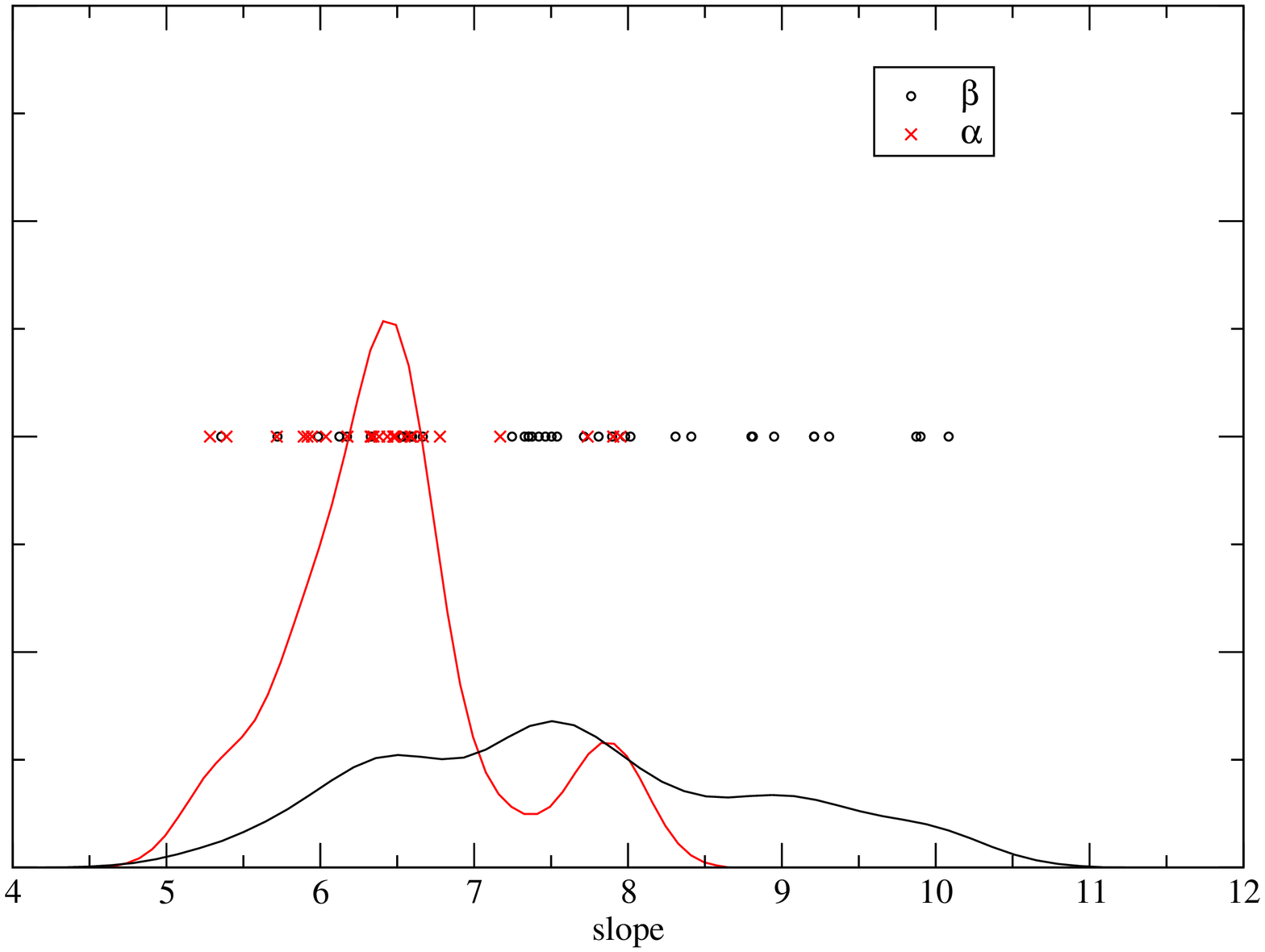}
  \caption{A scatter plot based on the slope (in time) of roughness for
the morphological classes alpha and beta. The density profile is also
shown.~\label{fig:rugoscatt}}
  \end{center}
\end{figure*}

With regard to the roughness of the aggregate surface,
Figure~\ref{fig:rugo} shows for a few typical cells the almost linear
increase in roughness with time up to 50 time units, which is chosen to
correspond to the deposition of 10 objects (cells). In order to
investigate the discriminative power of this measure, we produce a
scatter plot shown in Figure~\ref{fig:rugoscatt}, considering the slope
of the curves of roughness vs. time for all cells in our database. O que
esta figura mostra? O que estah na ordenada?  Table 1 indicates that a
Bayes classification analysis considering the roughness profile leads to
a reasonable classification rate for the alpha cell type (88
poorer discrimination of the beta cells (73
involves estimating the density functions of each class of a training
set of objects, weighted by the corresponding mass densities, and taking
as the more likely class of a new object the class presenting the
highest density for the specific set of measurements.

\begin{table}[ht]
\begin{center}
\begin{tabular}{l|c|c|c}\hline
         &  $\alpha$ &  $\beta$ & classification error\\\hline
$\alpha$ &  21 & 3  &  0.12\\\hline
$\beta$  &  10 & 26 &  0.27\\\hline
\end{tabular}
\end{center}
\caption{Bayes classification analysis considering the roughness profile
displayed in Figure~\ref{fig:rugoscatt}.~\label{table:rough}}
\end{table}

\begin{figure*}[htb]
  \begin{center}
    \includegraphics[scale=.45,angle=-90]{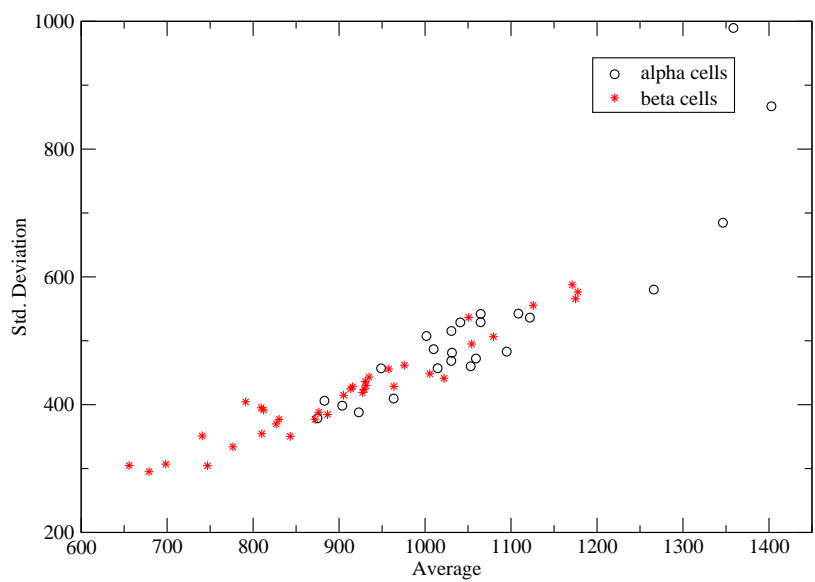}
\caption{A scatter plot defined by the mean and standard deviation
values of   the percolation index showing a reasonable separation of the
morphological cell types alpha and beta.~\label{fig:scatter}}
  \end{center}
\end{figure*}

\begin{figure*}[htb]
  \begin{center}
    \includegraphics[scale=.65,angle=0]{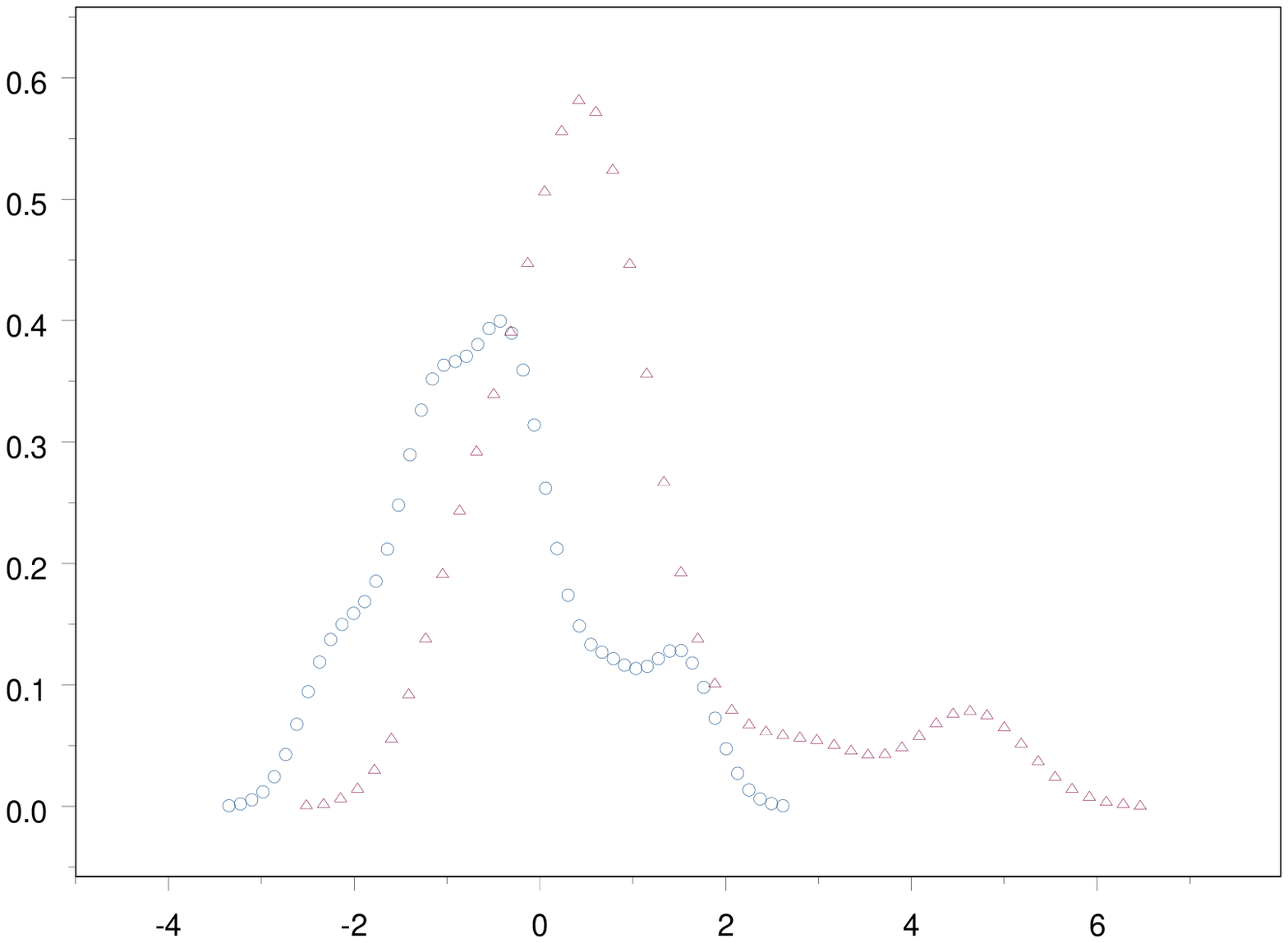}
  \caption{Density profile for the first principal component of the
measures defining the scatter plot in Figure~\ref{fig:scatter}, showing
clearly distinguishable peaks for the two morphological classes. The
curve with triangles represents the alpha cells and the other curve
represents the beta cells.~\label{fig:pcadensity}}2
  \end{center}
\end{figure*}

We now analyse whether the percolation index can also be used to
discrimate between alpha and beta cells. This is carried out by
extracting a series of global measurements such as the standard
deviation and the mean from the histograms in order to produce a feature
space or scatter plot.  Figure~\ref{fig:scatter} shows the scatter plots
for all cells in our database of retinal cells (53 cells). It is clear
from this scatter plot that while the retinal ganglion cells do share
similarities, reflecting the known difficulty for visual classification,
there is a clear division of the feature space populated by each kind of
cells. This is best illustrated by the density analysis shown in
Figure~\ref{fig:pcadensity}.  (o que estah nos eixos x e y da figura?)
In addition, analogously to the results for the artificial shapes, such
measurements are also highly correlated, suggesting that the correlation
between the mean and the standard deviation could be a more general
tendency. Table 2 shows the Bayesian classification considering the data
in Figure~\ref{fig:pcadensity}, indicating classification rates slightly
poorer than those obtained with the roughness measurements in Table 1.

\begin{table}[ht]
\begin{center}
\begin{tabular}{l|c|c|c}\hline
         &  $\alpha$ &  $\beta$ & classification error\\\hline
$\alpha$ & 20 & 4  &  0.16\\\hline
$\beta$  &  15 & 21 &  0.41\\\hline
\end{tabular}
\end{center}
\caption{The result of a Bayes classification analysis considering the
principal component profile displayed in
Figure~\ref{fig:pcadensity}~\label{table:perc}}.
\end{table}

\clearpage
\section{Conclusions}

The main aim of this work was to seek a means to quantify the potential
for contact between objects with complex shapes. Pursuing further the
suggestion to use the percolation critical density to quantify the
potential of connection between neuronal shapes, as reported
in~\cite{Costa:2003}, in the present work we addressed the critical
density characterizing the percolation induced by the contacts between
objects deposited in a ballistic process. While the percolation studies
reported in~\cite{Costa:2003} allowed overlap between parts of the
objects~\footnote{In the case of neuronal cells, such overlaps would
correspond to the fact that dendrites and/or axons in 3D spaces can
deviate one another in order to establish connections with any part of
the involved cells.}, the connections between objects in the present
investigations are limited to take place by contact induced by the
ballistic depositions.  Therefore, such connections tend to occur
between the most external parts of the objects.  An immediate
consequence is that the geometrical properties~\cite{Barbosa:2003} of
the interior of the cells become completely irrelevant to the critical
densities used in this article.  Thus, the percolation critical
densities obtained in the current work provide a quantification of the
potential of the objects to connect by contact, which is in principle
different from the values obtained in~\cite{Costa:2003}
and~\cite{Regina:cond-mat}.

In addition to the percolation critical densities, we also considered
the use of the roughness of the aggregate surface as a possible
measurement for characterizing neuronal morphology.  The potential of
both such features for discriminating between the two principal types of
ganglion neuronal cells (namely $\alpha$ and $\beta$ morphological
types) was also assessed in terms of Bayesian
classification~\cite{Costa:2001b}. The discriminating ability upon using
the percolation critical density or the roughness of the aggregate
surface can be objectively quantified in terms of the number of correct
classifications. In this context, the two measurements led to similar
performances, even though the roughness measurement had a slightly
superior discrimination between the two types of cell, as shown in
Figures~\ref{fig:rugoscatt} and \ref{fig:pcadensity} and
Tables~\ref{table:rough} and~\ref{table:perc}. Similarly to the results
obtained in the previous approach reported in~\cite{Regina:cond-mat},
the beta cells are more likely to engage in connections.  Such an
agreement between different percolation experiments, considering
complementary types of connectivity, could be understood as an
indication that the beta cells are optimized for enhanced connectivity.

Future investigations with the measurements considered in this article
may involve applications to 3D objects and comparison with other
measurements of shape morphology, such as shape
functionals~\cite{Barbosa:2003,Barbosa:2004,Barbosa:2004b} and
lacunarity~\cite{Rodrigues:2004}.

\bibliographystyle{unsrt}
\bibliography{choral}

\begin{thebibliography}{10}

\bibitem{Kandel:1996}
E.~R. Kandel, J.~H. Schwartz, and T.~M. Jessell.
\newblock {\em Principles of Neural Science}.
\newblock Appleton \& Lange, 1996.

\bibitem{Costa:1999}
L.~F. Costa and T.~J. Velte.
\newblock Automatic characterization and classification of ganglion cells from
  the salamander retina.
\newblock {\em Journal of Comparative Neurology}, 404:33--51, 1999.

\bibitem{Costa:2004}
L.~F. Costa, M.~S. Barbosa, and V.~Coupez.
\newblock A direct approach to neuronal connectivity.
\newblock {\em Physica A}, 341:618--628, 2004.

\bibitem{Barbosa:2003}
M.~S. Barbosa, L.~F. Costa, and E.~D. Bernardes.
\newblock Neuromorphometric characterization with shape functionals.
\newblock {\em Physical Review E}, 67(6), 2003.

\bibitem{Jones:2001}
C.~L. Jones and H.~F. Jelinek.
\newblock Wavelet packet fractal analysis of neuronal morphology.
\newblock {\em Methods}, 24(4):347--358, 2001.

\bibitem{Costa:2002}
L.~F. Costa, E.~T. Manoel, F.~Faucereau, J.~Chelly, J.~van Pelt, and
  G.~Ramakers.
\newblock A shape analysis framework for neuromorphometry.
\newblock {\em Journal of Comparative Neurology}, 13:282--310, 2002.

\bibitem{Barbosa:2004}
M.~S. Barbosa, L.~F. Costa, E.~S. Bernardes, G.~Ramakers, and J.~van Pelt.
\newblock Characterizing neuromorphologic alterations with additive shape
  functionals.
\newblock {\em The European Physical Journal B}, 37:109--115, 2004.

\bibitem{Costa:2001b}
L.~F. Costa and R.M. Cesar-Jr.
\newblock {\em Shape Analysis and Classification: Theory and Practice}.
\newblock CRC Press, 2001.

\bibitem{Montague:1991}
PR. Montague and MJ~Friedlander.
\newblock Morphogenesis and territorial coverage by isolated mammalian retinal
  ganglion-cells.
\newblock {\em Journal of Neuroscience}, 11(5):1440--1457, 1991.

\bibitem{Costa:2001}
L.~F. Costa, A.~G. Campos, and T.~M. Manoel.
\newblock An integrated approach to shape analysis: Results and perspectives.
\newblock In {\em QCAV2001}, pages 23--34, 2001.

\bibitem{Mandelbrot:1983}
Mandelbrot~B. B.
\newblock {\em The Fractal Geometry of Nature}.
\newblock W. H. Freeman, 1983.

\bibitem{Einstein:1998}
Einstein~A. J., Wu~H. S., and Gil J.
\newblock Self-affinity and lacunaritu of chromatin texture in benign and
  malignant breast epithelial cell nuclei.
\newblock {\em Phys. Rev. Lett.}, 80(2):397--400, 1998.

\bibitem{Rodrigues:2004}
Rodrigues~E. P., Barbosa~M. S., and Costa~L. da~F.
\newblock A self-referred approach do lacunarity.
\newblock {\em submited to Physical Review E}, 2004.

\bibitem{Costa:2003}
L.~F. Costa and E.~T. Manoel.
\newblock A percolation approach to neural morphometry and connectivity.
\newblock {\em Neuroinformatics}, 1(1):65--80, 2003.

\bibitem{Barabasi:1995}
A.~L. Barabasi and H.E. Stanley.
\newblock {\em Fractal Concepts in Surface Growth}.
\newblock University of Cambridge, 1995.

\bibitem{Sutherland:1966}
D.~N. Sutherland.
\newblock Comment on vold's simulation of floc formation.
\newblock {\em J. Colloid Interface Sci.}, 22:300--302, 1966.

\bibitem{Vold:1959}
M.~J. Vold.
\newblock Sediment volume and structure in dispersions of anisotropic
  particles.
\newblock {\em J. Phys. Chem.}, 63:1608--1612, 1959.

\bibitem{Regina:cond-mat}
Luciano da~F.~Costa and Regina~C. Coelho.
\newblock Growth-driven percolations: The dynamics of community formation in
  neuronal systems.
\newblock {\em q-bio.NC/0411009}, 2004.

\bibitem{Meakin:1988}
P.~Meakin.
\newblock Fractal aggregation.
\newblock {\em Adv. Colloid Interface Sci.}, 28:249--331, 1988.

\bibitem{Viot:1993}
P.~Viot, G.~Tarjus, and J.~Talbot.
\newblock Exact solution of a generalized ballistic-deposition model.
\newblock {\em Phys. Rev. E}, 48:480--488, 1993.

\bibitem{Cuisenaire}
O.~Cuisenaire and B.~Macq.
\newblock Fast euclidean distance transformation by propagation using multiple
  neighborhoods.
\newblock {\em Computer Vision and Image Understanding}, 76:163--172, 1999.

\bibitem{Barbosa:2004b}
L.~da~F.~Costa, M.~S. Barbosa, and V.~Coupez.
\newblock On the potential of the excluded volume and autocorrelation as
  neuromorphometric descriptors.
\newblock {\em Physica A}, Available online, 2005.

\end{thebibliography}
\end{document}